\documentclass[aps,pre,twocolumn,superscriptaddress,showpacs,floatfix]{revtex4-1}

\usepackage{graphicx}
\usepackage{amssymb,amsmath,amsbsy,amsfonts,bm}
\usepackage{epsfig}
\usepackage{subfigure}
\usepackage{multirow}
\usepackage{tikz}
\usepackage{hyperref}
\usepackage{gensymb}
\usepackage{threeparttable}
\usepackage{csquotes}
\usepackage{braket}
\usepackage{cleveref}
\usepackage[percent]{overpic}
\usepackage{color}

\DeclareFontFamily{OT1}{pzc}{}
\DeclareFontShape{OT1}{pzc}{m}{it}{<-> s * [1.10] pzcmi7t}{}
\DeclareMathAlphabet{\mathpzc}{OT1}{pzc}{m}{it}

\definecolor{darklavender}{rgb}{0.45, 0.31, 0.59}
\definecolor{red}{rgb}{1.0, 0.0, 0.0}
\definecolor{amethyst}{rgb}{0.6, 0.4, 0.8}
\definecolor{paulcolour}{rgb}{0.78, 0.082, 0.52}
\definecolor{changedcolour}{rgb}{1.0, 0.0, 0.0}
\definecolor{hlcolour}{rgb}{0.78, 0.2, 0.2}

\newcommand{\ofr}{(\mathbf{r})}

\newcommand{\figref}[1]{Fig.~\ref{#1}}
\newcommand{\secref}[1]{Sec.~\ref{#1}}
\newcommand{\equref}[1]{Eq.~(\ref{#1})}
\newcommand{\subfigref}[2]{Fig.~\ref{#1}.\textbf{(#2)}}

\newcommand{\degr}{\mathpzc{d}}

\newcommand{\nofr}{\mathbf{\hat{n}}\ofr}
\newcommand{\Lf}{L_\text{\text{f}}}
\newcommand{\Lb}{L_\text{\text{l}}}
\newcommand{\Qf}{Q_\text{\text{f}}}
\newcommand{\Qb}{Q_\text{\text{l}}}
\newcommand{\Qfb}{Q_\text{\text{f}}^{\text{(in)}}}
\newcommand{\Qbb}{Q_\text{\text{l}}^{\text{(in)}}}

\newcommand{\Qbbound}{Q_\text{l}^\text{(b)}}

\usepackage[normalem]{ulem}

\begin{document}
\title{Network topology of interlocked chiral particles}

\author{Paul A. Monderkamp}
\email{paul.monderkamp@hhu.de}
\affiliation{Institut f\"ur Theoretische Physik II: Weiche Materie, Heinrich-Heine-Universit\"at D\"usseldorf, 40225 D\"usseldorf, Germany}

\author{Rika S. Windisch}
\affiliation{Institut f\"ur Theoretische Physik II: Weiche Materie, Heinrich-Heine-Universit\"at D\"usseldorf, 40225 D\"usseldorf, Germany}

\author{Ren\'e Wittmann}
\affiliation{Institut f\"ur Theoretische Physik II: Weiche Materie, Heinrich-Heine-Universit\"at D\"usseldorf, 40225 D\"usseldorf, Germany}

\author{Hartmut L\"owen}
\affiliation{Institut f\"ur Theoretische Physik II: Weiche Materie, Heinrich-Heine-Universit\"at D\"usseldorf, 40225 D\"usseldorf, Germany}

\begin{abstract}

Self-assembly of chiral particles with an L-shape is explored by Monte-Carlo computer simulations in two spatial dimensions.
For sufficiently high packing densities in confinement, a carpet-like texture emerges due to the interlocking of L-shaped particles, resembling a distorted smectic liquid crystalline layer pattern.
From the positions of either of the two axes of the particles, two different types of layers can be extracted, which form distinct but complementary entangled networks.
These coarse-grained network structures are then analyzed from a topological point of view.
We propose a global charge conservation law by using an analogy to uniaxial smectics and
show that the individual network topology can be steered by both confinement and particle geometry. 
Our topological analysis provides a general classification framework for applications to other intertwined dual networks.

\end{abstract} 

\maketitle

\section{\label{sec_introduction}Introduction}
The response of any liquid crystal to external constraints, such as confinement, intricately depends on the density but crucially also on the geometry of the particles. In fluids of uniaxial rod-like particles, for instance, the particles tend to align at intermediate densities, forming a so-called nematic phase. At the same time, the rods favor certain alignments with confining walls, such that the
material becomes continuously deformed, to balance
the competition between these two factors \cite{yao2018topological,yaochen2020,rectangularConfinement,slitpores}.
Upon increasing packing fraction, the liquid crystal tends to transition into a smectic phase, where the particles additionally stack into layers. 
Consequently, confinement typically leads to a fragmentation into several domains, separated by grain boundaries \cite{annulus,basurto2020,monderkamp2021topology,monderkamp2022topological}.
\begin{figure}[h!]
\begin{center}
\includegraphics[width=1.0\linewidth]{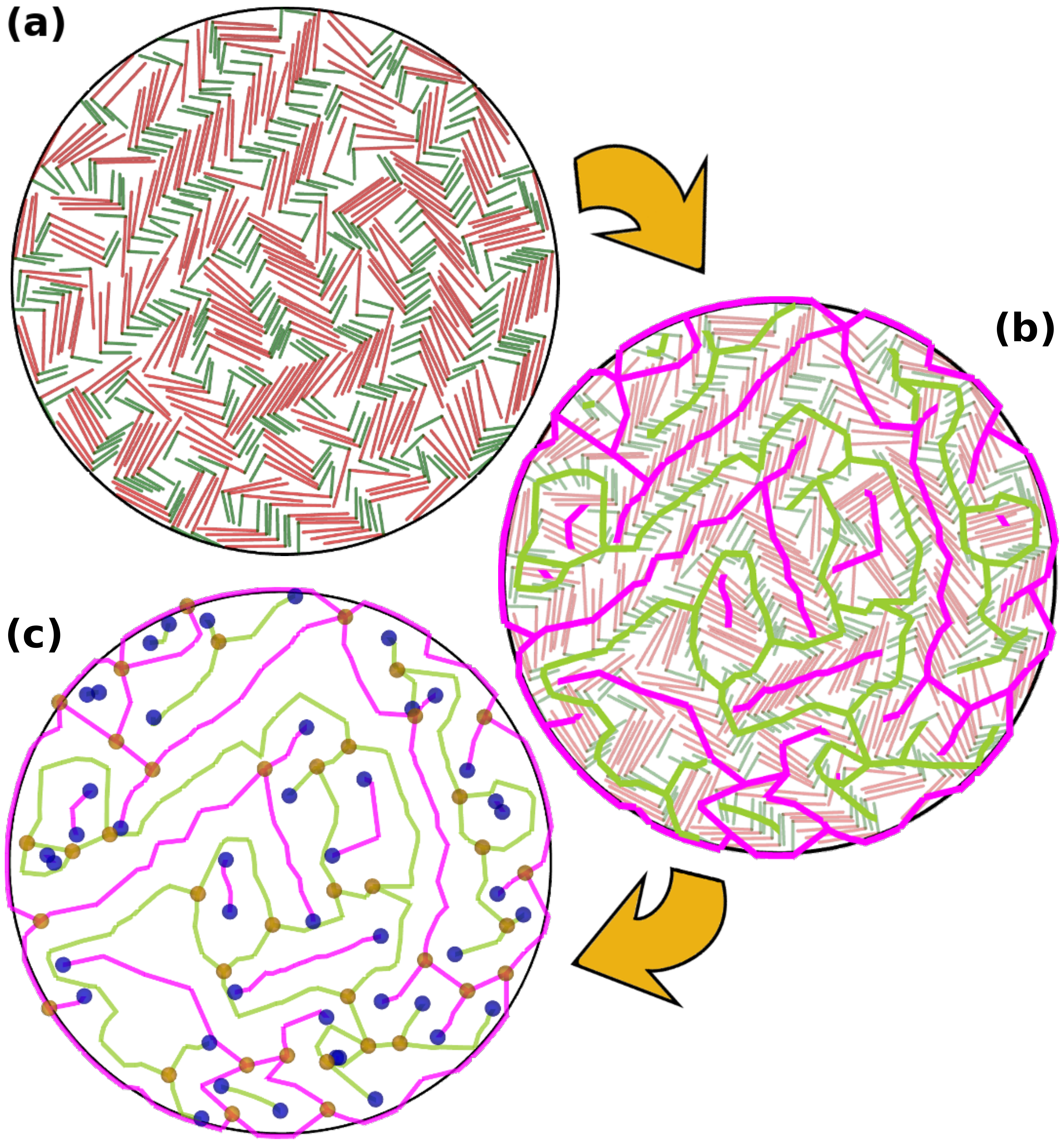}
\caption{\label{fig_network_concept}
Circularly confined L-particles on different levels of coarse-graining according to the yellow arrows. \textbf{(a)}: Simulation snapshot. The long axis of the L is colored in red, the short axis in green. \textbf{(b)}: The coarse-grained layer structures of both axes are visualized (long axis layers: magenta, short axis layers: green). \textbf{(c)}: Further coarse-grained structures with their vertices (dots). We assign the vertices a topological charge $q$, according to the number of adjacent edges (blue: $q>0$, yellow $q<0$). The number of these charges $q$ serves the quantification of disorder in the confined systems.
}
\end{center}
\end{figure}
Advances in the synthesis of molecular and colloidal liquid crystals, enables the study of systems with increasing geometrical complexity of the particles. Non-convex particles such as polygonal rings, banana-shaped particles and \textit{colloidal alphabet soups} allow for geometric interlocking, increasing the rigidity of the material \cite{avendano2016assembly,hernandez2007colloidal,niori1996distinct,heppke2000novel,dingemans2000non,ros2005banana}. 
In particular, particles with a characteristic chiral shape can exhibit interlocking at high densities providing an ideal playground for a wealth of interesting ordered structures \cite{kamien2001order,harris1997microscopic,harris1999molecular,pollard2019point,meyer1977ferroelectric,dierking2014chiral,hoell2016colloidal}.\\

In this work, we study the topology of smectic-like layers for interlocked chiral particles under the influence of confinement (see~\figref{fig_network_concept}).
To this end, with the aid of Monte-Carlo simulations, we generate systems of particles with the chiral shape of the letter L~(see~\figref{fig_particle_concept}), which are confined to circular and annular cavities. 
We observe the emergence of highly complex structures through the interlocking of the chiral, non-convex particles (see~\subfigref{fig_network_concept}{a}). 
In particular, both particle axes 
display a tendency for layering, while the competition between these two rigidly connected building blocks prevents the formation of regular layered patterns found in ordinary smectics. 
Each  layer associated with one axis of the L-shaped particles can be interpreted as a network such that the whole confined liquid crystal can be understood as an interwoven structure of two coexisting species of networks 
(see~\subfigref{fig_network_concept}{b}). We analyze the topology of the systems by only considering these coarse-grained networks and assigning an index to the vertices, depending on the number of adjacent edges (see~\subfigref{fig_network_concept}{c}).\\ 

The algorithm for the extraction of these layers is designed to create an output analogous to the topological picture of conventional uniaxial smectics. There, topological charge conservation is guaranteed by explicitly considering smectic layers (density peaks) and so-called half-layers in between (density minima) as topological entities, that carry topological charge \cite{machon2019,aharoni2017composite,beller2014geometry,hocking2022topological}. In that way, the indices of the vertices become topological, i.e., adhere to topological charge conservation in analogy to conventional electrodynamics, where the total charge, consisting of inside and boundary charge, adjusts to the topology of the confining container. In that way, the circle and the annulus yield different total topological charges, due to their different connectivity.\\

Moreover, we show that the structure of the individual networks can be largely steered by the particle shape. The vertical axis of the letter L is denoted as \textit{leg}, while the horizontal axis is denoted as \textit{foot}.
In particular, the spatial distribution as well as the amount of inside topological charges depends delicately on the \textit{foot} to \textit{leg} ratio $p = \Lf/\Lb$ (see Fig.~\ref{fig_particle_concept}) of the L-shaped particle. Specifically, the total amount of charge within the interior depends non-monotonically on $p$.\\

Finally, We expect that our general topological treatment of the convoluted network structures is also of practical value as a classification framework for other systems, where intertwined dual network structures can be found, such as gyroid phases in celestial nuclear matter \cite{schutrumpf2015appearance,nakazato2009gyroid}, technical applications \cite{crossland2009bicontinuous}, blood vessels in living organisms 
or traffic networks \cite{barthelemy2006optimal,papageorgiou1990dynamic}.\\

The paper is arranged as follows. In Sec.~\ref{sec_simulation}, we present our simulation protocol, while we elaborate on the graph-theoretical foundations of the network topology in Sec.~\ref{sec_nettop}. We present our results in Sec.~\ref{sec_results}, before we conclude in Sec.~\ref{sec_conclusion}.

\section{\label{sec_methods}Methods}
\subsection{\label{sec_simulation}Simulations}
\begin{figure}[t]
\begin{center}
\includegraphics[width=0.4\linewidth]{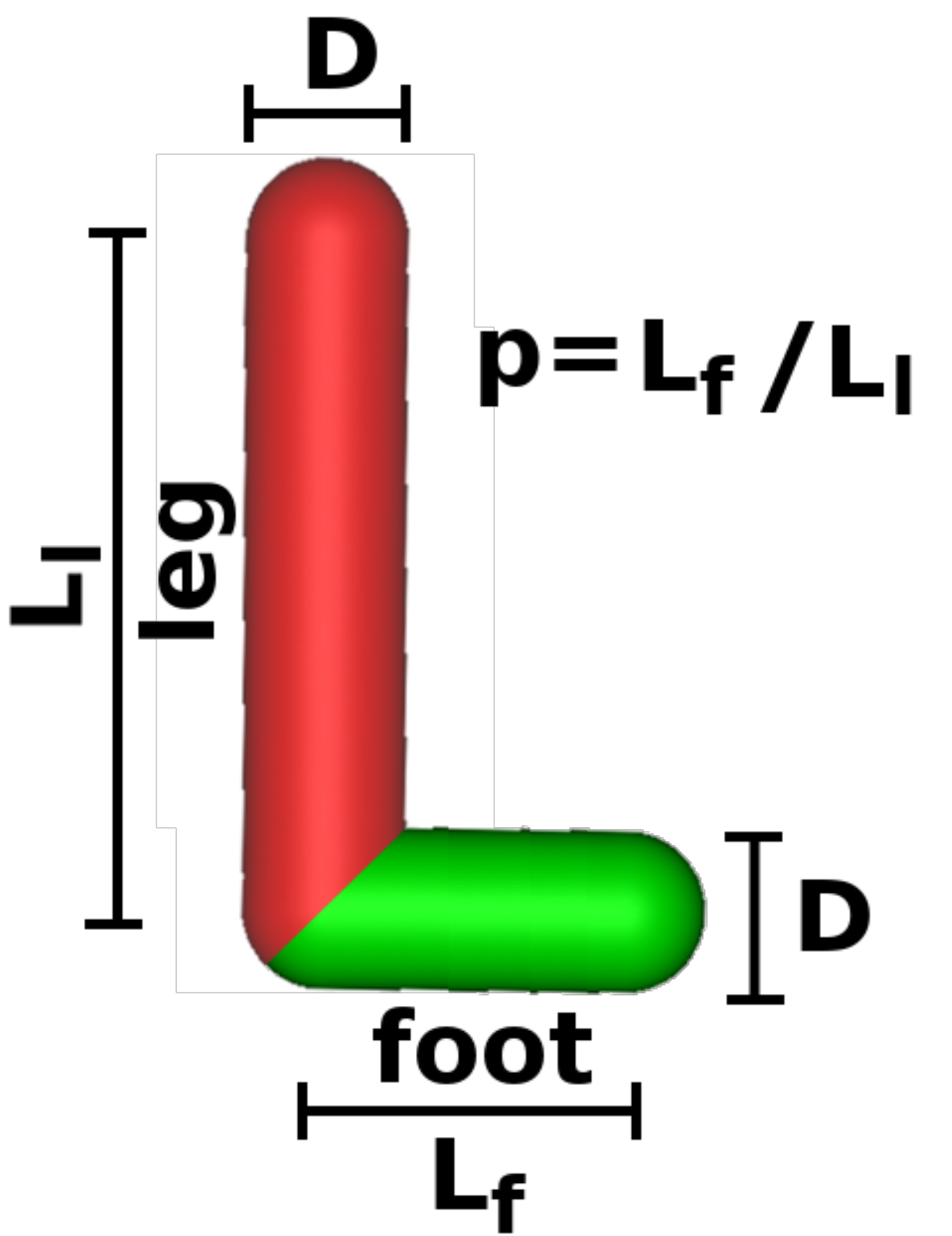}
\caption{\label{fig_particle_concept}
Schematic of the particles with the shape of the letter L as considered in this work. The lengths of the two axes are referred to as $\Lf$ (foot of the letter L, green) and $\Lb$ (leg of the letter L, red), respectively. The ratio of the axes is denoted by $p = \Lf/\Lb$.}
\end{center}
\end{figure}

We perform canonical Monte-Carlo simulations for particles that have the shape of the letter L (see \figref{fig_particle_concept}). We confine the particles to two-dimensional spherical and annular cavities.
The particles are modeled as a pair of rigidly connected almost hard discorectangles with equal diameters $D$, and core lengths $\Lf$, $\Lb$, expressed in units of $D$. Throughout this work, we vary $p = \Lf/\Lb$, while $\Lf+\Lb = 28D$ stays constant. The interactions of the L-shaped particles are conveniently specified through those of the individual rod-like building blocks.\\

For any two rods,  $i$ and $j$, not constituting to the same 
L-particle, with positions $\mathbf{r}_i$,  $\mathbf{r}_j$ and orientations $\mathbf{\hat{u}}_i$ and $\mathbf{\hat{u}}_j$, the pair potential is defined as harmonic repulsion \begin{equation}
\label{eq_Uijrod}
    U_{i,j} (\mathbf{r}_i,\mathbf{r}_j,\mathbf{\hat{u}}_i,\mathbf{\hat{u}}_j) =\left\{
\begin{array}{ll}
    U_0(\tau)  (D- d_{ij})^2 & \textnormal{ } d_{ij}<D \\
    0 & \textnormal{ }  d_{ij} \geq D,  \\
\end{array}
\right.
\end{equation}
\begin{equation}
d_{i,j} = \min_{\left | \alpha,\beta \right | < \frac{L}{2}} \left \| \mathbf{r}_i 
+ \alpha \mathbf{\hat{u}}_i - ( \mathbf{r}_j + \beta \mathbf{\hat{u}}_j ) \right \|,
\end{equation}
with $\alpha, \beta \in [-L/2,L/2]$, with $L$ in $\{\Lf,\Lb \}$ 
defines the overlap \cite{overl} and $U_0(\tau)$ is the interaction coefficient, which is linearly increasing as a function of the simulation progress $\tau \in [0,1]$, i.e., fraction of completed total of $10^6$ MC-cycles. 
$U_0(\tau = 1)$ is chosen as $10^3 k_\text{B} T$ to mimic almost hard repulsion in the equilibrated system.\\

We model the wall interaction by representing the rods as three virtual point-like particles sitting at the ends, and middle of the $k$-th rod $\mathbf{r}_\lambda = \mathbf{r}_k + \lambda \mathbf{\hat{u}}_k$, with $\lambda \in \{ -L/2 , 0 , L/2\}$. The interaction potential
\begin{equation}
\label{eq_wallpot}
   V(x)  =\left\{
\begin{array}{ll}
    V_0\left (\tau \right)\,x^2 & \textnormal{for } \mathbf{r}_\lambda \textnormal{ outside,}\\
    0 & \textnormal{for }  \mathbf{r}_\lambda \textnormal{ inside the cavity }\\
\end{array}
\right.
\end{equation}
with the walls is once again harmonic,
where $x$ denotes the closest distance of $\mathbf{r}_\lambda$ to the inside of the respective cavity. Similarly to $U_0$, $V_0$ is linearly increased as a function of the completed Monte-Carlo cycles $\tau \in [0,1]$ up to as $10^3 k_\text{B} T$ to mimic hard walls. We  simulate the systems at packing fractions $\eta_{\text{1}} = Na_\text{L}/a_\text{cav} = 0.4$ and $0.6$, with the area of a single L-shaped particle $a_\text{L}$ and the area of the cavity $a_\text{cav}$. The radii of the confinements are typically in the range $ 4L \lesssim R \lesssim 50L$. Correspondingly, typical particle numbers $N$ are between several hundreds and several thousands.
In the annular confinement, we keep the width of the annular ring constant ($\Delta R \approx 1.83\Lb$), while we vary the inner radius ($R_\text{in}$) and outer radius ($R_\text{in} + \Delta R$).\\

We follow a compression protocol, where we randomly initialize the system at a low volume fraction $\eta_{\text{0}} = 0.1 \eta_{\text{1}}$. Each MC-cycle consists of a trial displacement or rotation of each particle.
The acceptance probability $P=\min(1,\exp(-\Delta U/k_\text{B} T))$, for any trial move, is given by the Metropolis criterion from the difference $\Delta U$ of the energies (see Eqs.~(\ref{eq_Uijrod}) and (\ref{eq_wallpot})) in the system before and after trial \cite{metropolis}. Over the course of the simulation, we gradually rescale the positions of the walls and particles to increase $\eta$ according to
\begin{equation}
\label{eq_compression}
    \eta (\tau) =
 \eta (\tau) = (\eta_{\text{1}} - \eta_{\text{0}}) {\tau} ^{\frac{1}{3}} + \eta_{\text{0}  }
\end{equation}
until the volume fraction $\eta_{\text{1}} = 0.4$ is reached.
This type of decelerating compression aids the equilibration speed, since the system is quickly compressed in the dilute stage, while being allowed to undergo a larger fraction of MC-cycles in the regime, where self-assembly of the ordered structures is expected
(for more details on the protocol, see Appendix~\ref{sec_app_protocol}).

\subsection{\label{sec_nettop} Network topological charge analysis}

As illustrated in \figref{fig_particle_concept}, 
we denote the vertical (red) axis by \textit{leg} which we distinguish from the horizontal \textit{foot} (green) axis by presuming a parallel wall alignment of the \textit{leg} (see~Appendix~\ref{sec_app_netgen} for more details). 
As visible in \figref{fig_network_concept}, each ensemble of smectic-like layers associated with either \textit{foot} or \textit{leg} (perceived as individual rods)
forms convoluted networks and correspondingly can be analyzed with the help of mathematical graph theory.\\

Each network consists of a set of vertices . Those vertices are connected via a set of edge lines \cite{euler1758elementa,alama2008euler}. 
\begin{figure}[t]
\begin{center}
\includegraphics[width=1.0\linewidth]{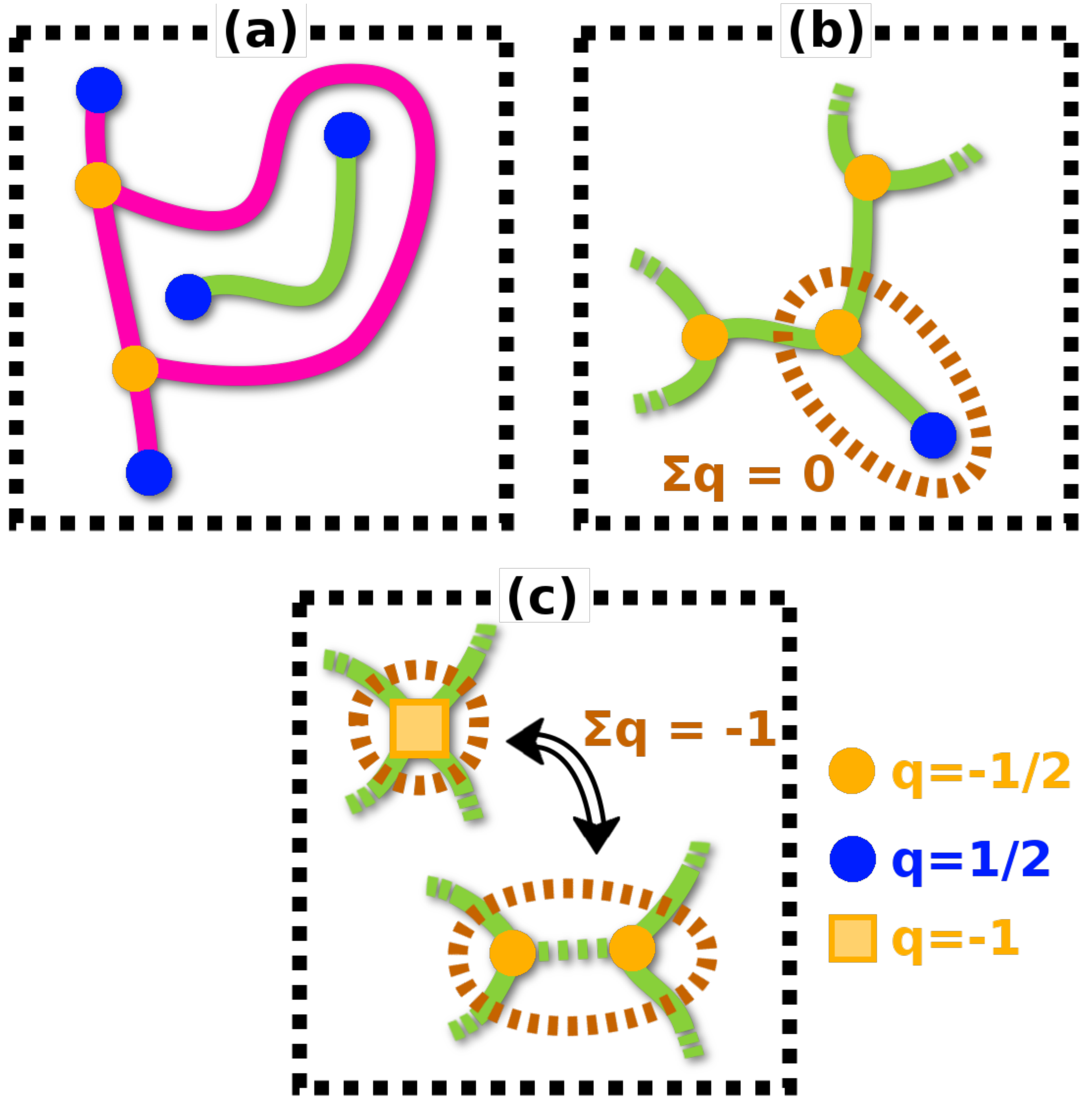}
\caption{\label{fig_nettop} Schematic of the network topological model (see Eq.~\eqref{eq_netwQ}).
\textbf{(a)}: Serving charge conservation, our algorithm is designed such that there can not be empty loops. As such, any loop contains a network of the respective other species. The total network topological charge in the system is conserved under this condition, since the charges of a loop and a simply connected network cancel. 
\textbf{(b)}: The total charge $q$, within any area of a single network, is calculated from the number of in-/outgoing edges through the boundary, inserted into Eq.~\eqref{eq_netwQ}. Dangling ends, such as within the dashed circle, are charge neutral. Vertices with two adjacent edges, carry no charge, and are therefore not explicitly labeled.
Non charge-neutral operations are \textit{(i)} adding loops ($\Delta q = -1$) or (ii) new networks ($\Delta q = +1$) adds to the total charge within the system. 
\textbf{(c)}: A vertex with four outgoing edges ($\degr=4$, $q=-1$ see Eq. \ref{eq_netwQ}) can be understood as two infinitesimally close vertices with $\degr=3$, $q=-1/2$, respectively.
} 
\end{center}
\end{figure}
As known from the treatment of the topology of layers in conventional smectics \cite{machon2019,aharoni2017composite,beller2014geometry,hocking2022topological}, a charge conservation follows, if the species of networks alternate. In other words, between any two smectic layers has to be a density minimum, i.e., a half-layer. Accordingly, our graph generation is designed, such that empty loops are contracted into a single vertex (see Appendix~\ref{sec_app_netgen}).
Therefore, we only observe the occurrence of loops (see \figref{fig_network_concept}), that each envelop a simply connected graph (without loops) of the other species. This is schematically visualized in \figref{fig_nettop}.\textbf{(a)}.
More seldom, multiple loops of alternating species are nestled into each other, with a simply connected graph in the middle. To characterize this behavior in general, we assign a topological network charge to any vertex in the network as 
\begin{equation}
   \label{eq_netwQ} 
   q = 1 - \frac{\degr}{2}
\end{equation}
with the degree $\degr$, i.e., the adjacent number of edges.
Note, that this definition of the network charges is analogous to the layer topological charges, typically considered in smectic liquid crystals, where the edge lines represent smectic layers (see Appendix~\ref{sec_app_convsmec}).
As such, the total charge of a network species reads as
\begin{equation}
    Q_a = \sum_{\substack{\text{vertices}\\\text{in $a$}}} q
\end{equation}
with $a\in\{ \text{f}, \text{l}\}$, standing for \textit{foot} and \textit{leg}. where $\Qf + \Qb$ is a conserved quantity.

\begin{figure*}[t]
\begin{center}
\includegraphics[width=0.9\linewidth]{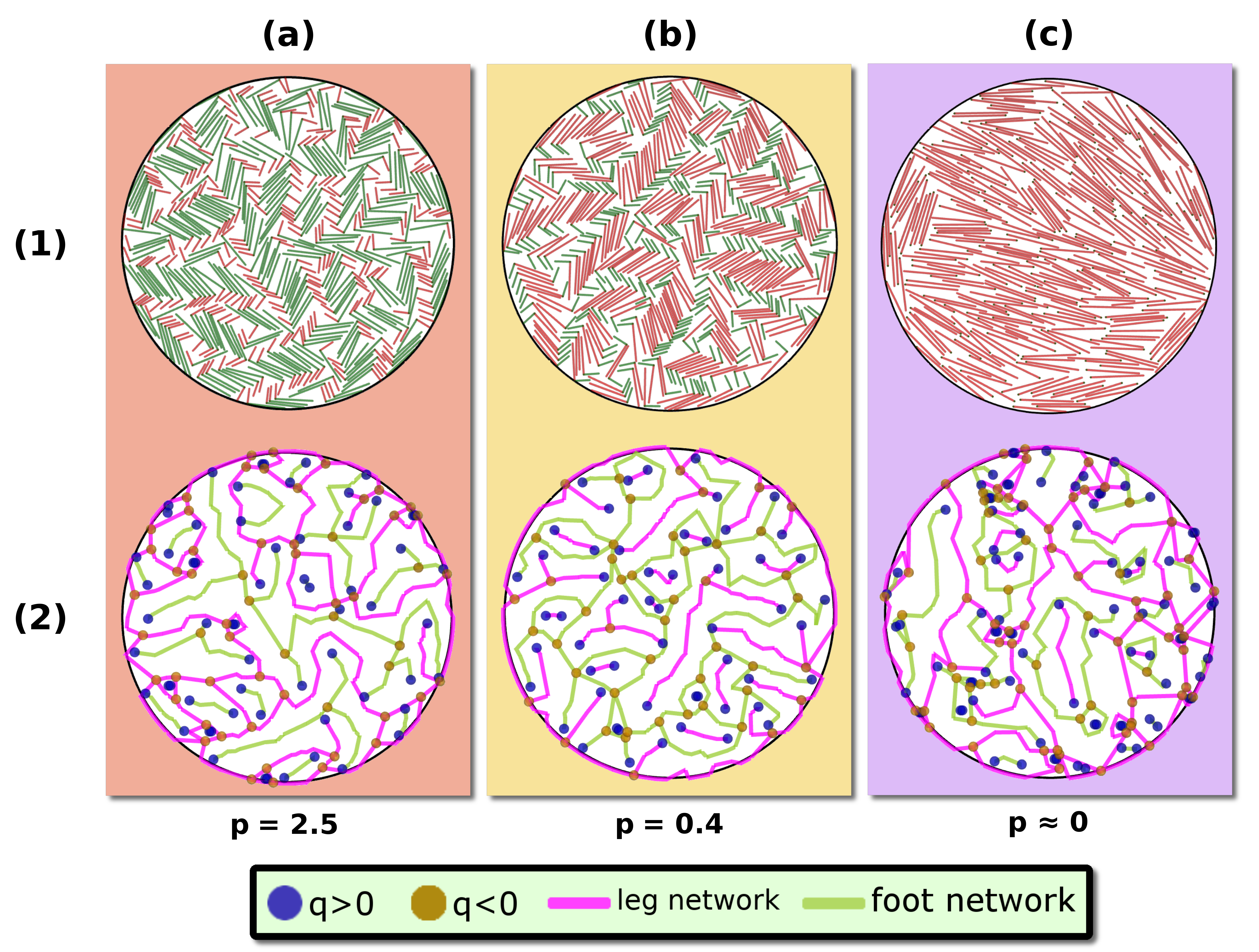}
\caption{\label{fig_circlesnaps} Results of three independent simulation runs for L-shaped particles confined to spherical cavities with packing fraction $\eta = 0.4$ and \textbf{(a)}: $p = 0.4$, \textbf{(b)}: $p = 2.5$ and the hard-rod limit \textbf{(c)}: $p \approx 0$. The upper row \textbf{(1)} displays particle snapshots superimposed with the layer networks of both species (see sec. \ref{sec_nettop}). The lower row \textbf{(2)} shows the isolated networks. The negative network charges are indicated in brown, while the positive are shown in blue. In all cases, the boundary charges are assigned to the magenta \textit{leg}-network.
Therefore the inside charge of the \textit{foot}-species, matches the respective total charge ($\Qfb = \Qf$), which is not true for the \textit{leg}-species ($\Qbb \neq \Qb$).
Accordingly (a) and (b) correspond to a color swap, only in the inside of the cavity.
The total network charge in the system $\Qf + \Qb$ matches the Euler characteristic of the confinement $\chi = 1$.
} 
\end{center}
\end{figure*}

The total charge of a network is only determined by its connectivity and not by the total number of vertices.
As visible from \figref{fig_nettop}.\textbf{(b)}, adding a dangling end to a previously existing network, is a charge neutral operation. Consequently, any simply connected network, carries the charge of a single isolated vertex $q = 1$.
Only the addition of loops, i.e., adding an edge between two existing vertices, decreases the net charge by $1$. 
This network topological charge can only be conserved, if every loop coincides with a simply connected (loop-less) network, e.g., at its center.\\

This definition of the charge is consistent, such that the charge within any area, can be calculated from the number of in-/outgoing edges, similar to Gauss's law in classical electrodynamics (see \figref{fig_nettop}.\textbf{(c)}). 
Any vertex of higher degree can be viewed as a composition of merged $q=-1/2$ charges.
Vertices with $\degr=2$ carry no charge, and can be therefore, together with the edges, viewed as constituting the layers, in between the rest of the charges.
Respecting these properties of the network charge, our algorithm which generates the final networks (see Appendix~\ref{sec_app_netgen}) is designed to systematically eliminate vertices, such that the final network structure, as shown in \subfigref{fig_network_concept}{c}, both illustrates the network connectivity and allows us to properly determine the total charges $\Qf$ and $\Qb$.\\ 

Typically, considering the conservation of topological charges in confined geometries, e.g., orientational topological defects in nematically ordered fluids \cite{real_defch,monderkamp2021topology}, requires the identification of boundary defects on the system walls if no global alignment condition is obeyed. 
Due to the invariance of the total charge within the cavity $Q=\Qf+\Qb$ (determined by the topology of the confining domain),
we have the liberty to assign the outer walls of the confinement to any of the two network species. In the following, without loss of generality, we choose to assign the boundary defects to the layer network of the \textit{leg}, generating the boundary charge $\Qbbound$. 
We will denote the inside network charges, i.e., charges without explicit inclusion of boundary charge, as $\Qfb$ and $\Qbb$. As detailed in Sec.~\ref{sec_results}, the total sum 
\begin{equation}
\Qfb+\Qbb =Q-\Qbbound
\end{equation}
is not conserved. Still, these quantities contain structural information about the confined state.
Through our choice of assigning the boundary charges to the \textit{leg}-network, clearly $\Qfb = \Qf$ but $\Qbb \neq \Qb =\Qbb+\Qbbound$.

\section{\label{sec_results}Results}

\subsection{Circular confinement \label{sec_circonf}}

For circular cavities, we show simulation results for systems of L-shaped particles, with three different ratios of axes lengths $p$ in \figref{fig_circlesnaps}.
at  packing fraction $\eta = 0.4$.
In \subfigref{fig_circlesnaps}{a1} a typical snapshot, for $p = 2.5$, is depicted. 
The particles locally interlock, while the longer \textit{foot}-axes display a strong tendency for alignment, leading to elongated clusters. Global orientational ordering, however, is not visible, such as would be expected from, e.g., a conventional smectic liquid crystal. 
Through the interlocking, the shorter \textit{leg}-axes fill the spaces in between. Below, in \subfigref{fig_circlesnaps}{a2}, we show the corresponding graph network, resulting from our analysis of the layers. The elongated \textit{foot}-clusters are represented in \subfigref{fig_circlesnaps}{a2} by the green network. The \textit{leg}-network is depicted in magenta. The relative higher stiffness of the wider \textit{foot}-layers results in a favoring of longer simply-connected networks, each contained in a loop of a single large magenta \textit{leg}-network. This results in the presence of majorly negative defects (indicated in brown), due to the loops. Conversely, every separated component of the \textit{foot}-network contains a charge $\sum q = 1$. As elaborated in \secref{sec_nettop}, this results in a total network charge within the system of $Q=1$. Similarly, \subfigref{fig_circlesnaps}{b} shows a simulation snapshot for the inverse length ratio $p = 0.4$. 
Accordingly, the behavior of the networks within the interior of the confinement, visible in \subfigref{fig_circlesnaps}{b2} is analogous to the former case with inverted species. 
The assignment of the boundary charges remains with the \textit{leg}-network, as in \subfigref{fig_circlesnaps}{a}. As the total charge is invariant of our choice of this alignment condition, the inclusion of the boundary charges still retains total network charge within the confined system as $Q=1$. Finally, \subfigref{fig_circlesnaps}{c} shows a simulation snapshot, where $p \approx 0$, i.e., hard rods. The confined system resides in a state, where global orientational ordering is present. Through the lack of interlocking, no strong indications of layers are visible, as expected from the unconfined nematic bulk phase which is stable at the chosen packing fraction $\eta=0.4$ \cite{phase_beh_DF,bolhuis1997tracing}. Nevertheless, the network analysis can still be used to quantify the global topology. Naturally, the total charge is still conserved as $Q=1$.\\

\begin{figure}[t]
\begin{center}
\includegraphics[width=1.0\linewidth]{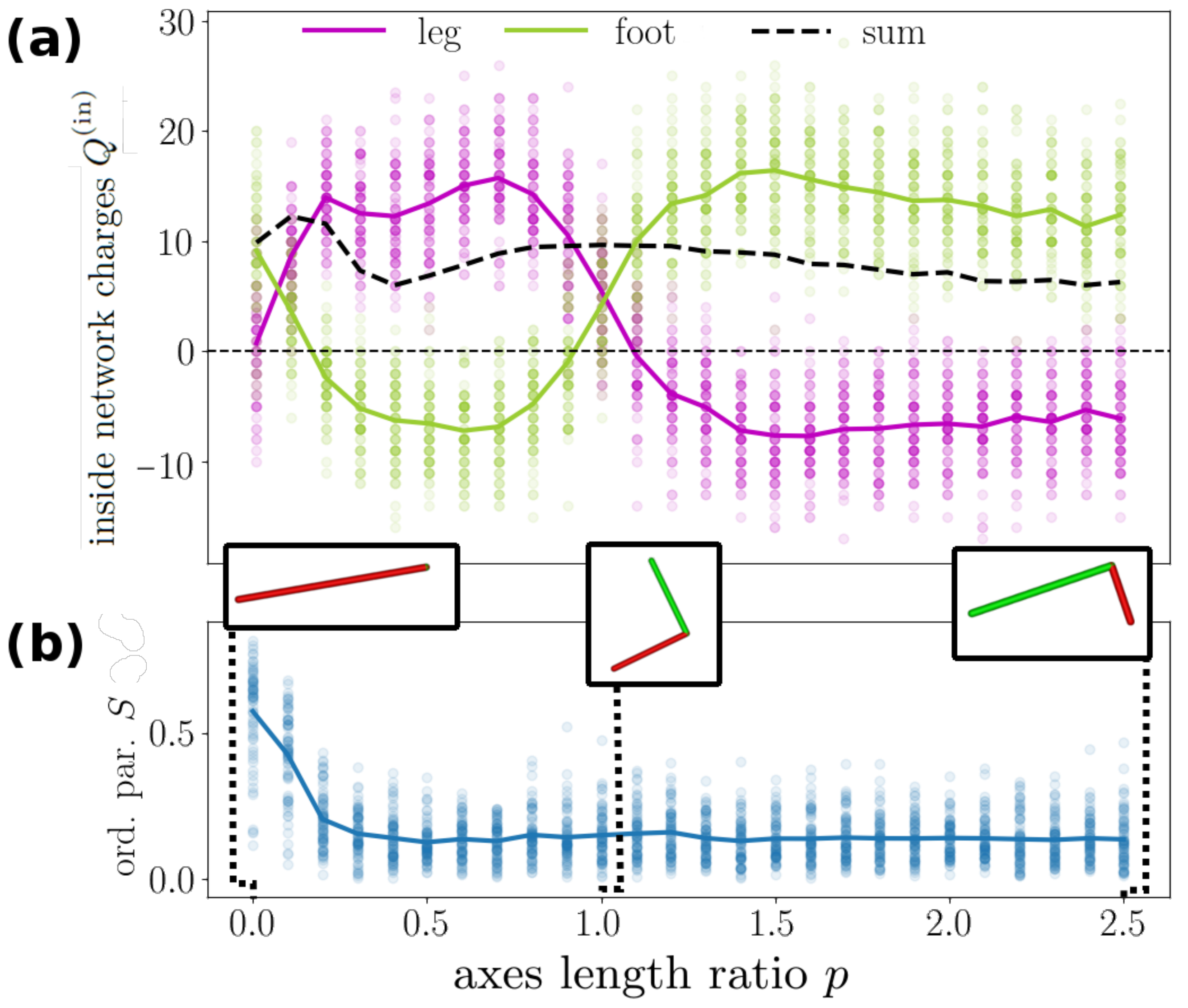}
\caption{\label{fig_Qofrat}
Simulation results of liquid crystals composed of L-shaped particles for a range of different length ratios $p$ of both particle axes, illustrating a topological transition in the networks at $p=1$, where the chirality of the particles flips. \textbf{(a)}: Inside charges $\Qfb$ (green), $\Qbb$  (magenta) and sum $\Qfb+\Qbb$ (dashed black) 
\textbf{(b)}: Global orientational order parameter $S$.  
We observe, that at each point of the horizontal axes, the respective longer axis network has a positive charge, and the shorter axis network a negative charge. Above approximately $0.2$, the systems reside in interlocked layered states without global orientational order. Below $0.2$, the systems converge against conventional nematic order.
The results are shown at packing fraction $\eta = 0.4$ .The lines correspond to averages over $75$ simulations per $p$.\\
}
\end{center}
\end{figure}

In \subfigref{fig_Qofrat}{a} we show the inside charges $\Qfb$ and $\Qbb$ of both network species as well as their sum, i.e., the total charge without the inclusion of boundary charges, as a function of $p$. Additionally, along the same horizontal axis, in \subfigref{fig_Qofrat}{b}, we show the global orientational order parameter $S  = | \left < \exp(i2\phi) \right > |$, where  $\left < ... \right >$ denotes an average over all particles. The data shown in both figures can be roughly divided into three characteristic regions along the horizontal axis, namely $p > 1$ (cf.~\subfigref{fig_circlesnaps}{a}), $ 0.2 \lesssim p < 1$ (cf.~\subfigref{fig_circlesnaps}{b}, as well as $p \lesssim 0.2$ (cf.~\subfigref{fig_circlesnaps}{c}). In the case $p > 1$, the simply-connected, but isolated \textit{foot}-layers (green) result in positive inside charge. At the same time, the single large \textit{leg}-network with loops results in negative charges (magenta). Analogously, the same holds true in the regime $ 0.2 \lesssim p < 1$, only with inverted network species. Therefore, the signs of $\Qfb$ and $\Qbb$ are flipped.
For $p \lesssim 0.2$, the shapes of the particles are approaching the hard-rod limit.
The absence of an entropic contribution from the interlocking mechanism of the L-shaped particles, allows the liquid crystal to undergo a transition into a nematic state with global orientational ordering. Correspondingly, \subfigref{fig_Qofrat}{b} shows an increased orientational order parameter $S$. In the transition regime, complicated packing effects dominate the system, causing non-trivial behavior in $\Qfb$ and $\Qbb$. 
We observe that the first peak in $\Qbb$ decreases for larger systems, where the interior of the confinement is less influenced by the system boundaries (not shown). 
We therefore infer, that the behavior in \figref{fig_Qofrat} stems from the extreme influence exerted by the confinement. 
Due to the symmetry of the particles, all the observables are symmetric around $1$, i.e., $S(p) = S(1/p)$, as well as $\Qbb(p) = \Qfb(1/p)$ (see Appendix~\ref{sec_app_Lratlonger}). Naturally, the total charges $\Qf$ and $\Qb$ do not adhere to this symmetry, which is broken by the assignment of the boundary charges to the \textit{leg}-network.
\\

In general, the total inside charge $\Qfb + \Qbb$ within the cavity, i.e., the sum of all network charges in \figref{fig_Qofrat} is  constantly greater than zero. This is consistent with our observation, of the presence of isolated networks within the inside, without the existence of empty loops.
Moreover, including the boundary to the \textit{leg}-network does by construction only add negative defects, i.e., $\Qbbound<0$, compare, e.g., \subfigref{fig_circlesnaps}{2}.
Taking a closer look at the behavior of $\Qfb+\Qbb$ in \subfigref{fig_Qofrat}{a}, we notice two local maxima.
The first one at $p\approx0.1$ coincides with the onset of global orientational order, i.e., close to the transition between confined nematic and interlocked layer states.
The second one is located around $p=1$, i.e., where neither network dominates the structure, such that the mutual interruption of the layering of the two building blocks is most pronounced.
We thus conclude that the total inside charge $\Qfb+\Qbb$ provides a useful measure for disorder in interlocked or frustrated systems.

\subsection{Annular confinement \label{sec_annconf}}

\begin{figure}[t]
\begin{center}
\includegraphics[width=0.9\linewidth]{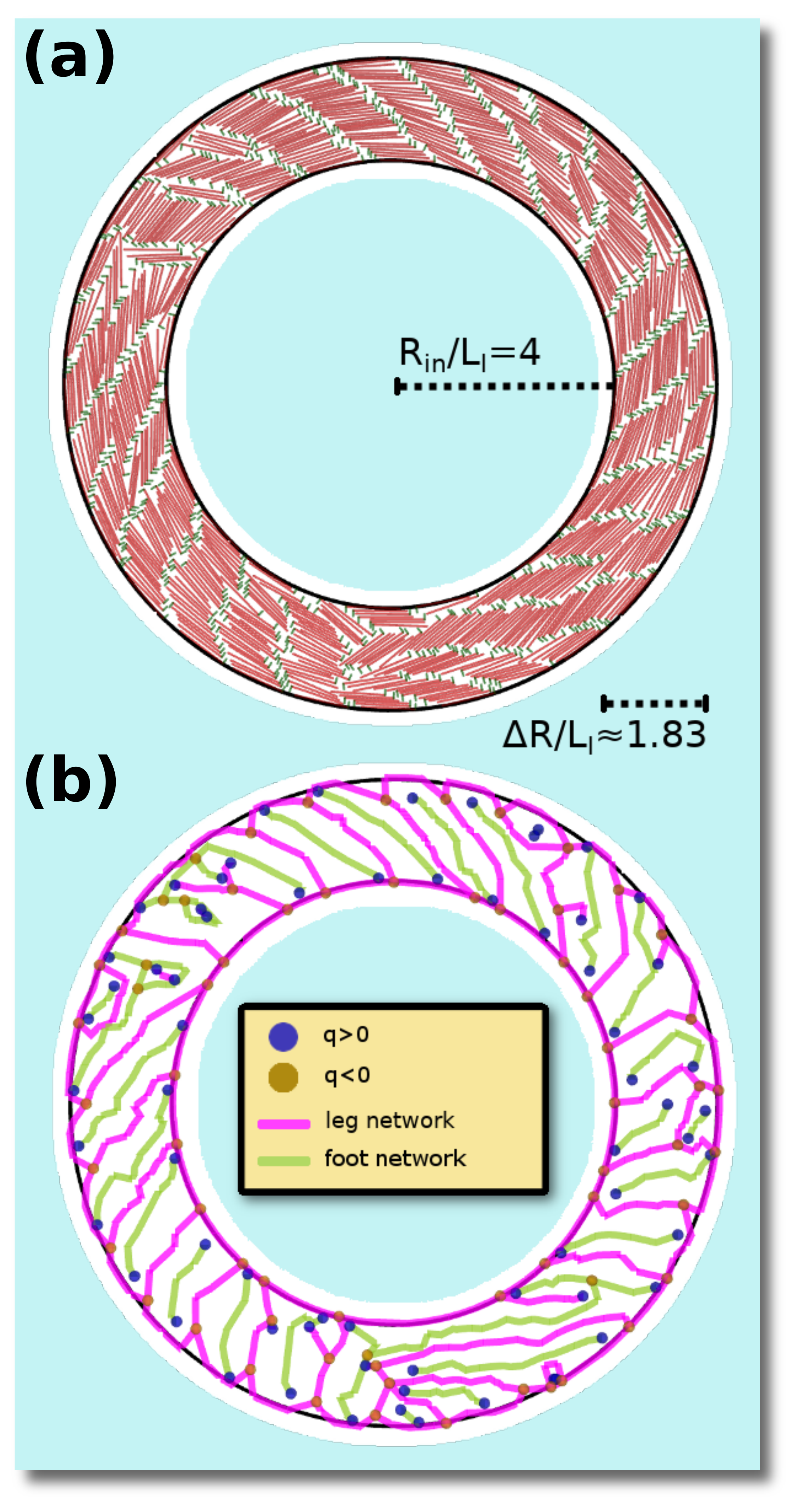}
\caption{\label{fig_annulus_snaps}
Simulation results for a system of L-shaped particles in annular confinement. \textbf{(a)}: The system resides in a smectic-$C$-like state, where due to the interlocking, the layer director is at an angle to the local orientation. \textbf{(b)}: The network structure roughly shows one \textit{foot}-network per smectic layer. The boundary charges are assigned to the \textit{leg}-network, therefore the network charge is conserved. Through the empty loop on the inner wall, the total network charge matches the Euler characteristic $\chi = 0$ of the confinement.
Packing fraction and axes length ratio are $\eta = 0.6$ and $p = 0.1$.}
\end{center}
\end{figure}

\begin{figure}[t]
\begin{center}
\includegraphics[width=1.0\linewidth]{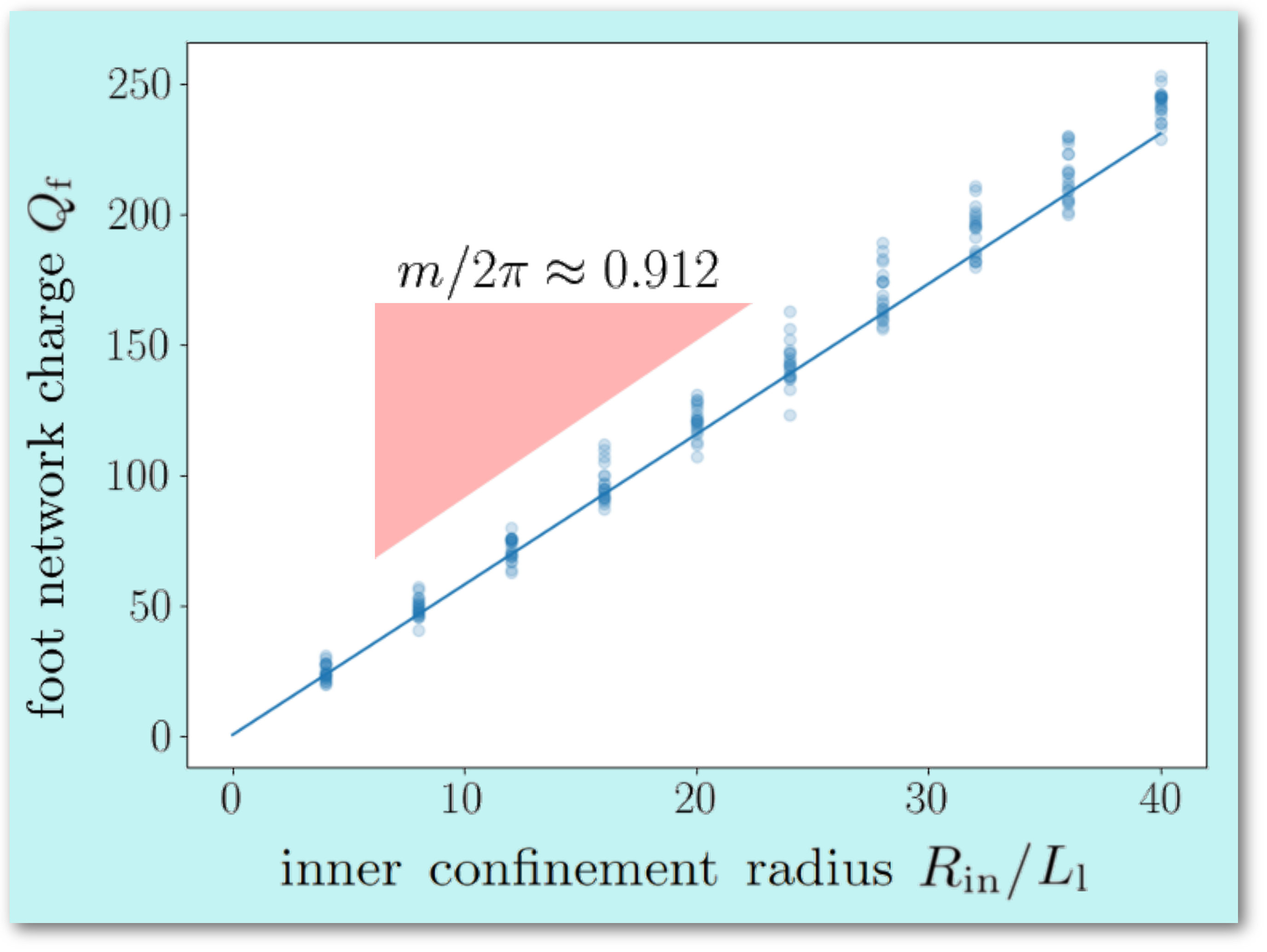}
\caption{\label{fig_Rinnerser}
Network topological charge of the \textit{foot}-species $\Qf =-\Qb$ within annular confinement as a function of the inner radius $R_\text{in}$. Packing fraction and particle dimensions match \figref{fig_annulus_snaps}.
We show results of $20$ simulations per considered $R_\text{in}$ and a linear fit.
The slope of the line $m$ divided by $2\pi$ is approximately one, confirming that there is slightly less than one charge per particle length around the perimeter. This represents, as expected, one smectic-like layer per particle length (cf. \figref{fig_annulus_snaps}). 
}
\end{center}
\end{figure}

One of the important characteristics of a topological conservation law within a confined liquid crystal system is the relevance of the topology of the confining container \cite{algebTop,nem_defects}. To further explore this, we introduce a confining domain with annular shape, that possesses 
Euler characteristic $\chi=0$.
In \subfigref{fig_annulus_snaps}{a}, a corresponding particle snapshot is shown ($R_\text{in} = 4\Lb$, $\Delta R \approx 1.83\Lb$, $\eta = 0.6$, $p = 0.1$). Through the high packing fraction and the relatively long \textit{leg}-axes (red), the particles show a visible tendency to reside in locally parallel layers, where the long axes of the particles are parallel align with the outer walls. At the same time, the protruding \textit{foot}-axes (green) cause a relative shift of neighboring particles, resulting in a characteristic smectic-$C$-like shape, where the particles are tilted with respect to the layers. 
The associated network structure is visible in \subfigref{fig_annulus_snaps}{b}. As in the previous section, the boundary charges are assigned to the \textit{leg}-network. Each smectic block of particles results in an isolated simply connected \textit{foot}-network, wrapped into a loop of the large \textit{leg}-network. The inner confinement walls cause an additional empty loop. As a result, all charges sum up to the Euler characteristic of an annulus $\chi = Q =  0$.\\

\begin{figure}[t]
\begin{center}
\includegraphics[width=1.0\linewidth]{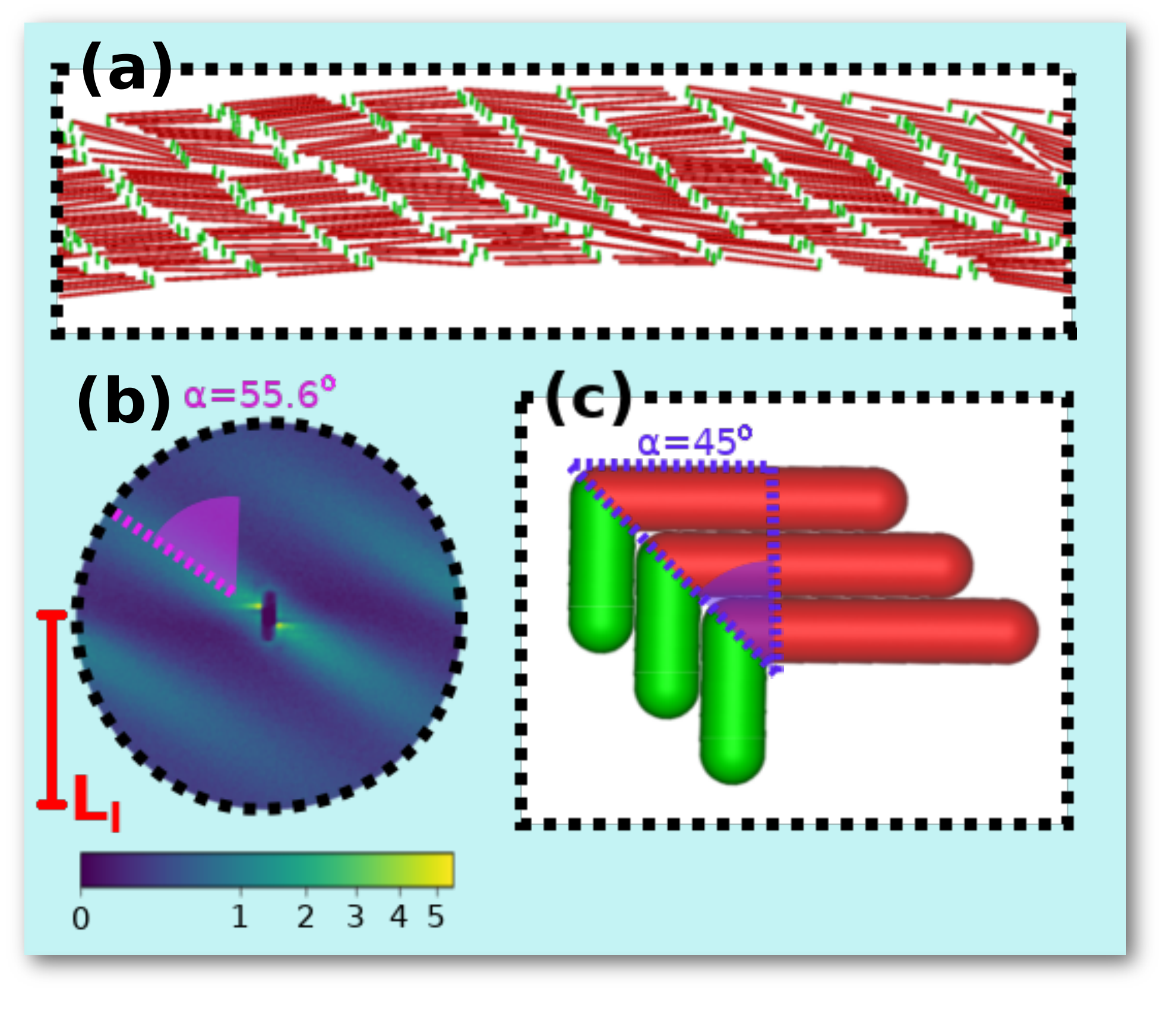}
\caption{\label{fig_annangles}
Simulation results in annuli with large inner radii $R_\text{in} = 40\Lb$. The other simulation parameters match \figref{fig_annulus_snaps}. \textbf{(a)}: Excerpt from a simulation snapshot. The tilt angle of the layers with respect to the walls of the annular ring is clearly visible. \textbf{(b)}: Annular pair distribution function $g_\circ (\mathbf{r})$ between the positions of the \textit{foot}-particles (green) in the $xy$-plane up to a radial distance of $\Lb$ (cf.~\equref{eq_gofr_ann}).
The slanted peaks represent the slanted \textit{foot}-layers in (a). The angle relative to the horizontal is equal to $\approx 34.4^\circ$.
\textbf{(c)}: schematic of L-particles in close packing. 
For $D_\text{foot} = D_\text{leg} = D$, the optimal tilt angles of the layers is equal to $45^\circ$ (vertical and horizontal sides of the right triangle in the figure have to be equal). The deviation in the measured angle in (b) possibly stems from complex entropic interactions of the layer species at lower packing fraction. 
}
\end{center}
\end{figure}

The previous results are for systems with relatively small particle numbers $N=320$.
We make this particular choice to illustrate the network charges on the particle-resolved level. Naturally, all of the above holds true also for larger systems. 
In Fig.~\ref{fig_Rinnerser}, we present the total topological charge $\Qf = \Qfb= -\Qb$ of the \textit{foot}-species as a function of the inner radius $R_\text{in}$ of the annular confinement, while keeping the width of the annulus $\Delta R = 1.83\Lb$ constant. 
Since the boundary charges are assigned to the \textit{leg}-network, the total charge $\Qf$ coincides with the inside charge $\Qfb$.
Each depicted point represents an individual simulation. 
As visible in the plot, the data points lie fairly accurately on a straight line, indicated by a linear fit with slope $m = 0.912 \times 2 \pi$. More specifically, there is slightly less than one positive charge per particle length along the inner circumference of the annulus. This is consistent with the observations made in \figref{fig_annulus_snaps}, where each particle layer forms a new network. Since any particle layer has a width, which is slightly larger than $\Lb$, $m/(2\pi)$ is slightly smaller than one.\\

To further investigate the origin of the smectic-$C$-like tilted layers for these L-shaped particles, additional results are presented in \figref{fig_annangles}. In conventional hard-rod smectics, confined to annular cavities, the layers typically align with the outer walls, while the direction of the layer is typically in radial direction of the confinement, i.e., orthogonal to the walls \cite{annulus,yaochen2020}. Here, we observe layers which are tilted with respect to the radial direction.
This is nicely visible in \figref{fig_annulus_snaps} and \subfigref{fig_annangles}{a}.
The latter shows an excerpt from a simulation in an annulus with inner radius $R_\text{in} = 40\Lb$. We see similar tilt angles in both snapshots, irrespective of the large difference in curvature of the respective confinement walls. In order to show the local structure within the annular ring, we introduce the two-dimensional annular pair distribution function
\begin{align}
\label{eq_gofr_ann}
    g_\circ ( \mathbf{r}  ) = \frac{1}{N\rho} \left < \sum_{\substack{i,j=1 \\ i\neq j}}^N \delta(\mathbf{r} - \overline{\overline{\mathbf{R}}} \cdot  \left ( \mathbf{r}_i - \mathbf{r}_j \right) )\right >.
\end{align}
This is obtained via the matrix 
\begin{align}
\overline{\overline{\mathbf{R}}} = \overline{\overline{\mathbf{R}}}(\phi_i) = \left( \begin{array}{rr}
\cos ( \pi/2 - \phi_i) & -\sin ( \pi/2 - \phi_i)  \\ 
\sin ( \pi/2 - \phi_i) & \cos ( \pi/2 - \phi_i)  \\
\end{array}\right),
\end{align}
where $\phi_i$ is the polar angle of the position $\mathbf{r}_i$ of the $i$-th particle with respect to the annulus center.
Here, $N$ and $\rho$ are the global particle numbers and densities, respectively, and $\delta(\mathbf{r})$ denotes the delta-distribution.
The physical interpretation of $g_\circ(\mathbf{r})$ is a distribution function of rotated relative vectors.
The data to obtain $g_\circ(\mathbf{r})$ are sampled from $20$ independent simulation runs.
We compute $g_\circ(\mathbf{r})$ for the positions of the green \textit{foot}-particles, as depicted in \subfigref{fig_annangles}{b}. In the depiction, the horizontal axis denotes the tangential direction of the annular walls.
The center of the diagram shows an anisotropic depletion zone around the particles, resulting from the almost hard repulsion. Furthermore, slanted density peaks are visible, representing the tilted layers within the annular ring. These peaks form an approximate angle of $\alpha = 55.6^\circ$ with the radial direction. In \subfigref{fig_annangles}{c}, it is illustrated that for identical width of the particle axes ($D_\text{foot} = D_\text{leg} = D$), an angle of $\alpha = 45^\circ$ leads to efficient packing at extreme packing fractions. At lower packing fractions, however, this would lead to a drastic decrease of the free length out of the layer, parallel to the longer \textit{leg}-particles. More specifically, the measured angle can be understood as a result of two competing entropic factors: \textit{(i)} the aforementioned free length along the \textit{leg} orientation, which tends to increase the tilt angle relative to the radial direction, and \textit{(ii)} the systems tendency to reside in a smectic-like layered structure perpendicular to the \textit{leg} orientation, due to the high density, which tends to decrease the tilt angle (as presumably, the layered structure breaks apart for $\alpha \rightarrow \pi/2$).

\section{\label{sec_conclusion}Conclusion}

In this work, we present a formalism for the analysis of the topology of liquid crystals with chiral particle shapes, which give rise to irregular defect structures in confinement. This approach is based on the consideration of the entangled network structure formed by the respective axes positions. 
In order to analyze the structures with two layer species, we generate networks utilizing Delaunay triangulations and identify topological charges, from the degrees (number of adjacent edges) of the network vertices. This procedure leans on the concept of layers and half-layers that characterize the topology in conventional smectics \cite{machon2019,aharoni2017composite}, as, e.g., formed by uniaxial hard rods. 
Like these previous applications, our study relies on coexisting but disjoint network species, which as a whole adhere to a common topological conservation law, where the sum of the respective topological charges in a confined system equals the Euler characteristic $\chi$ of the finite container. 
However, there are two crucial differences. \textit{(i)} In our study, both network species explicitly refer to an axis of the particles.
In contrast, in conventional smectics, one layer species represents the density minima in between the particle layers. \textit{(ii)} Our approach is based on the analysis of the layer network structure on the particle-resolved level, while the observation of conventional smectic point defects (see~Appendix~\ref{sec_app_convsmec}) typically happens on much larger length scales and was hitherto
majorly employed in approaches that describe the coarse-grained smectic layer structure \cite{machon2019,aharoni2017composite,beller2014geometry,hocking2022topological,stannarius2016defect,harth2020topological}.
Moreover, we exemplify that the bare inside charges serve as a useful quantifier of disorder.\\

We use  this framework of network topology to analyze the structure of liquid crystals composed of particles with the shape of the letter L, confined to circular ($\chi =1$) and annular ($\chi =0$) cavities. These are generated via the use of canonical Monte-Carlo simulation. We find that the global inside charges of the two network species intricately depend on both the local particle symmetries and the global degree of order. We observe a variety of remarkable states at different packing fractions and L-axes lengths: at intermediate densities, the particles prefer interlocking with suppression of global orientational order, when both axes of the L-shape have comparable sizes. Otherwise the system tends towards a conventional nematic state \cite{yao2018topological,yao2021defect,phase_beh_DF}. 
At high packing fractions in the annulus, the particles arrange in more rigid layers, which resembles a smectic-$C$ structure. In regards to the latter, we present additional simulation for annuli with large radii elucidating the origin of the tilt angle of the layers.\\

Based on this insight, we expect that the formalism used throughout this work can positively contribute to a variety of future endeavors. In particular we expect, that it will be useful in the interpretation of future computational, theoretical and experimental studies of systems with complicated particle shapes \cite{martinez2022effect,zhao2007nematic,zhao2012local,avendano2017packing,barmes2003computer,kraft2013brownian,yuan2018chiral,lapointe2009shape,zerrouki2008chiral}. Even though we introduce this method as a tool for the investigation of relatively complex chiral particles, we also suspect that it will be insightful to apply it to hard-rod fluids, since the network analysis can be readily applied to visualize the half-layers in conventional smectics. We expect it to provide also insight, applied in the analysis of gyroid phases in systems of particles without chirality \cite{schonhofer2017purely,schonhofer2018double}. This will possibly be a powerful tool in understanding the emergent topological structures \cite{bramble2007,FocalDomains,liarte2015}.
Recent years have also seen an increased interest in non-equilibrium systems, where the individual particles consume energy from their surrounding to propel themselves \cite{tan2019topological,loewe2021passive,giomi2015geometry,REVbaerSPR2020}. It has been shown, that in these systems, the orientational defects have dynamical properties, which in turn depend on the respective topological charges \cite{decamp2015orientational,hydroOfDef,huang2022defect}.
It seems therefore very reasonable to assume, that network topological charges will carry interesting dynamical properties in collectively moving L-shaped or other chiral swimmers \cite{hernandez2020collective,kummel2013circular,caprini2019active,lowen2016chirality}.

\section*{Acknowledgements}
The authors would like to thank Michael te Vrugt and Raphael Wittkowski for stimulating discussions.
This work is funded by the Deutsche Forschungsgemeinschaft (DFG, German Research Foundation) -- LO 418/20-2.

\section*{Author declarations}

\subsection*{Conflict of Interest}

The authors have no conflicts to disclose.

\subsection*{Author contributions}

\textbf{Paul A. Monderkamp}:
Conceptualization (equal);
Data Curation (lead);
Formal Analysis (lead);
Methodology (equal);
Project Administration (equal);
Software (equal);
Supervision (equal);
Validation (lead);
Visualization (lead);
Writing/Original Draft Preparation (lead);
Writing/Review \& Editing (equal);
\textbf{Rika S. Windisch}:
Conceptualization (equal);
Data Curation (supporting);
Formal Analysis (supporting);
Methodology (equal);
Project Administration (supporting);
Software (equal);
\textbf{Ren\'e Wittmann}:
Conceptualization (equal);
Formal Analysis (supporting);
Funding Acquisition (supporting);
Methodology (equal);
Project Administration (equal);
Supervision (equal);
Writing/Review \& Editing (equal);
\textbf{Hartmut L\"owen}:
Conceptualization (equal);
Funding Acquisition (lead);
Methodology (equal);
Project Administration (equal);
Resources (lead);
Supervision (equal);
Writing/Review \& Editing (equal);

\subsection*{Data availability}

The data that support the findings of this study are available
from the corresponding author upon reasonable request.

\appendix

\begin{figure}[t]
\begin{center}
\vspace{0.5cm}
\includegraphics[width=0.85\linewidth]{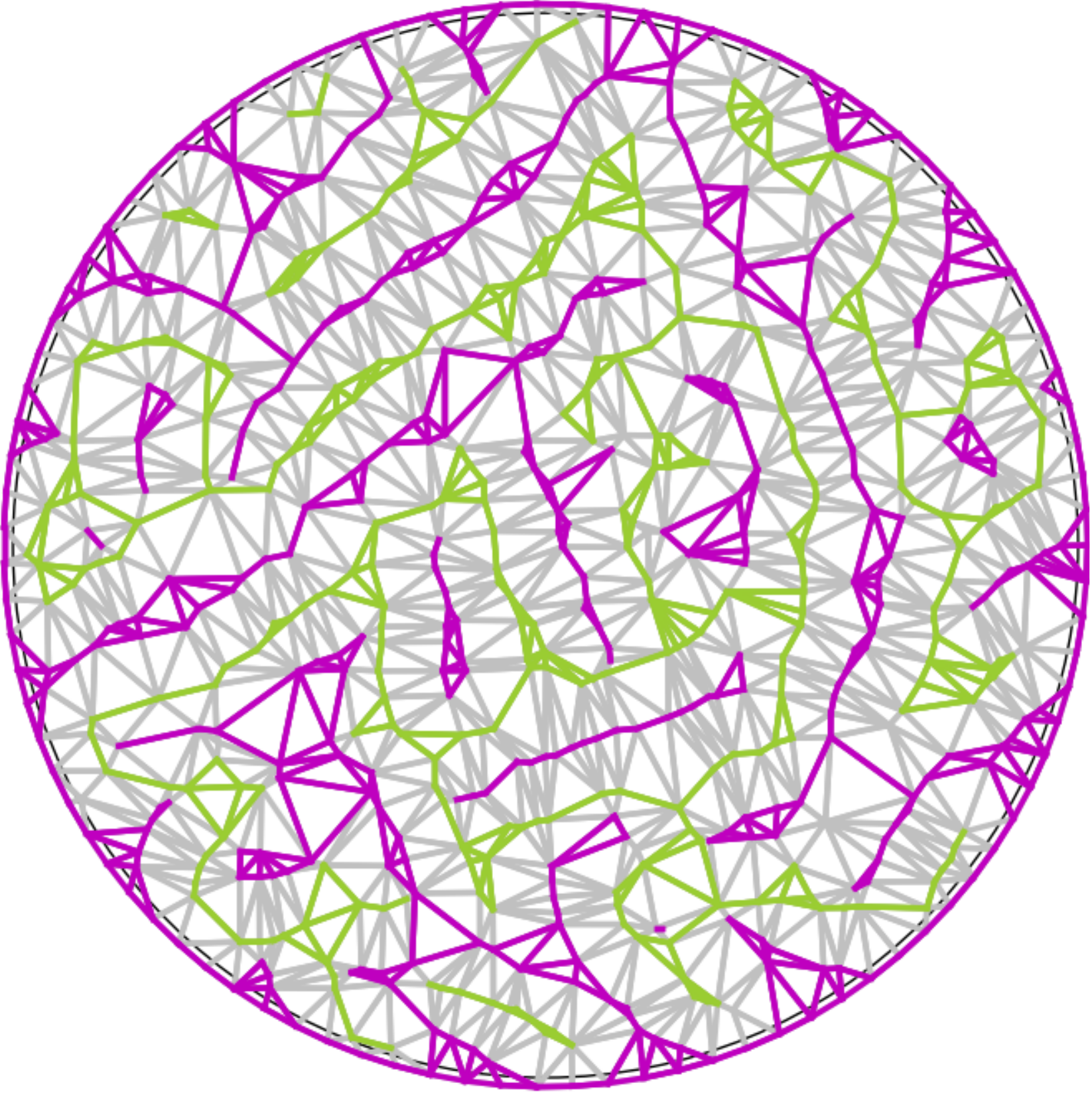}
\caption{\label{fig_app_graphgen}
Delaunay triangulation of the union of all coordinate positions of both axes of all L-shaped particles, used as a base point for the generation of the graphs networks, considered in this manuscript. The corresponding particle snapshot is shown in \figref{fig_network_concept}. The edges, connecting the \textit{foot}-positions are colored in green, while the edges, connecting the \textit{leg}-axes are colored magenta. Grey edges connect \textit{foot}- and \textit{leg}-vertices. In order to obtain lines, representing the layer structure of the respective axes, the gray edges are deleted, and the remaining colored networks are transformed into simple lines.} 
\end{center}
\end{figure}

\section{Details on the equilibration protocol \label{sec_app_protocol}}
In the main manuscript, the simulation results are generated with the help of a canonical Monte-Carlo simulation, as described in \secref{sec_simulation}. The fundamental goal of the procedure is to find a configuration at a relatively high packing fraction which reflects the equilibrium state. Since it is practically impossible to obtain a randomized configuration at the target packing fraction (since guessing a configuration which fits into the cavity corresponds to guessing the final simulation result), the systems are initialized several orders of magnitude below the target packing fraction. They are subsequently compressed with a decelerating compression rate (cf.~Eq.~\eqref{eq_compression}). Additionally, the interaction constant $U_0$ (cf.~Eq.~\eqref{eq_Uijrod}) between the particles and the wall interaction constant $V_0$ (cf.~Eq.~\eqref{eq_wallpot}) are linearly increased as a function of the fraction of completed Monte-Carlo cycles $\tau \in [0,1]$.
The initial softness of the interactions helps the particles heal-out overlaps, which may occur in the beginning of the simulation due to random initialization. At the end of the simulation, $U_0$ and $V_0$ are $10^3 k_\text{B} T$.
We find that the $\tau$-dependence of these constants becomes less relevant, as the simulation progresses, since the particles effectively feature hard repulsion at a certain point. This positively contributes to the equilibration.

To obtain configurations, which reflect the equilibrium configuration, without explicitly evaluating free energies, the equilibration is performed slowly enough, such that two assumptions can be made about the transient configurations over the course of the simulation. \textit{(i)}: The system is ergodic (samples the whole configuration space), in the stage of self-assembly, such that the system configuration space at higher densities is also sampled fairly over independent different simulation runs. \textit{(ii)}: The compression of the system and increase of the interaction strengths occurs slowly enough, that the system can equilibrate faster than the parameters change, such that the final configuration reflects 
a state close to equilibrium.

Below, we show additional simulation results, which aim to support both assumptions. The simulations are performed as described in \secref{sec_simulation} with axis length ratio $p=1$ and fraction $\eta_1 = 0.4$ inserted into the formula for compression (\equref{eq_compression}).
But instead of equilibrating up to $\eta_1 = 0.4$, the compression and increase of the interaction constants are stopped at the half-point of the simulation $\tau =1/2$. This value coincides with $\eta \approx 0.32$ (cf.~\equref{eq_compression}) Subsequently, the simulation runs at constant parameters for the second half to illustrate the change of the system without compression at intermediate densities.

\begin{figure}[t]
\begin{center}
\includegraphics[width=1.0\linewidth]{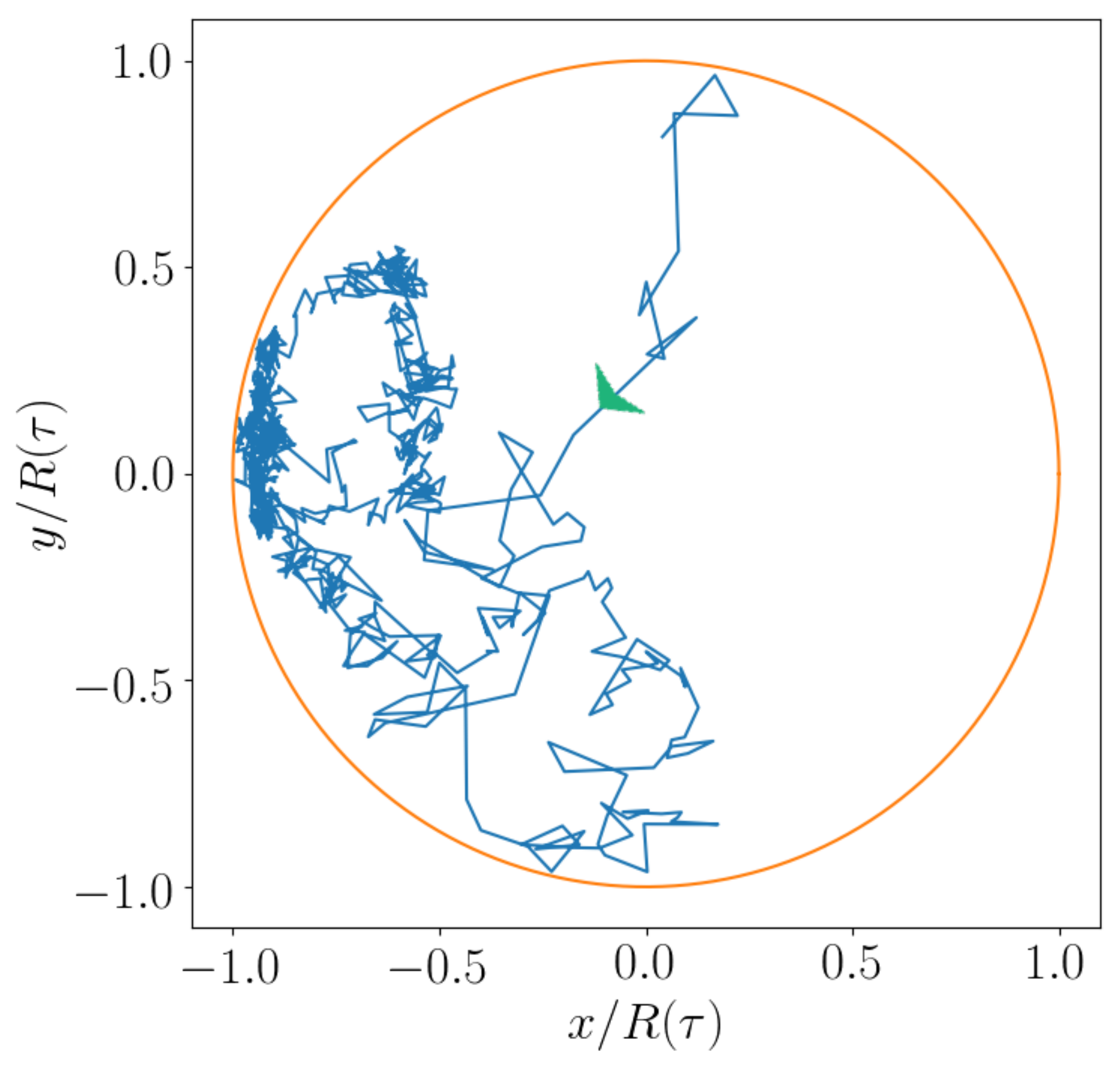}
\caption{\label{fig_particle_traj}
Trajectory of a single particle within a Monte-Carlo simulation, where the compression and change of interaction constants is stopped at half the number of total Monte-Carlo cycles $\tau = 1/2$. 
The simulation starts as described in \secref{sec_simulation}, but runs with constant parameters after $\tau = 1/2$.
The displayed particle positions are rescaled with the radius of the confinement, i.e. the depicted trajectory shows $\mathbf{r}(\tau)/R(\tau)$. 
The green arrow denotes the direction of motion within the trajectory. Note that the trajectory reflects Monte-Carlo displacement (contrary to a standard equation of motion). 
The relatively unconstrained motion suggests, that the system is in principle able to occupy all configurations. The particle also samples the whole orientation space (not shown).}
\end{center}
\end{figure}

\subsection{Ergodicity}

Here, we support the assumption, that the simulation protocol samples the configuration space ergodically at intermediate densities. This is to guarantee that the configuration space at higher densities is sampled fairly across many simulations. To this end, we show the trajectory of the \textit{leg}-axis of a single particle over the course of the simulation in \figref{fig_particle_traj}. The displayed positions are rescaled with radius of the confinement $R(\tau)$ ($R(\tau)^2 \propto 1/\eta(\tau)^{1/2}$, cf.~\equref{eq_compression}). As such, the figure shows $\mathbf{r} (\tau)/R(\tau)$. 
The fact that the particle moves throughout the whole cavity strongly suggests, that the entire configuration space is sampled fairly, since the particles can freely rearrange at intermediate densities, and therefore the system is not hindered from occupying specific configurations. Note that not every single particle has to be able to traverse the whole cavity, since they can be considered physically indistinguishable in terms of the states. 
We consider this numerical indication, that the system arrests into a high-density state, which is close to the equilibrium and independent of the initialization.

\begin{figure}[t]
\begin{center}
\includegraphics[width=1.0\linewidth]{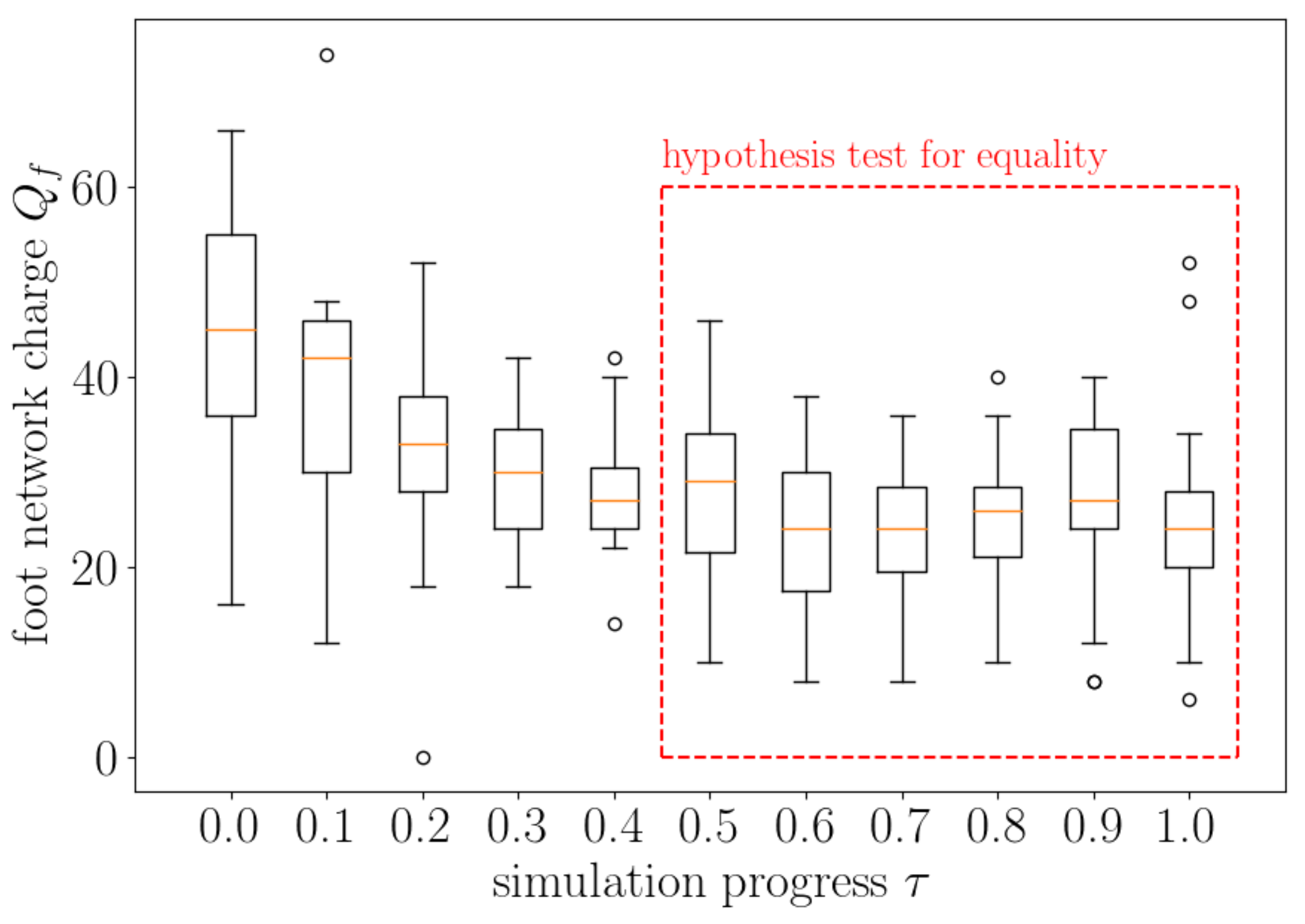}
\caption{\label{fig_Qoftau}
Total network charge of the \textit{foot}-network as a function of simulation progress $\tau$, i.e., the fraction of completed Monte-Carlo cycles. The compression and adjustment of interaction constants is stopped at the half-way point, after which the simulation runs at constant parameters. 
We pick six instances of simulation progress at which we average $\Qf$. Those averages are denoted $\overline{\mathcal{Q}}_f$ (see red box).
We perform a hypothesis test for equality of all six $\overline{\mathcal{Q}}_f$, via a one-way $F$-test. We find an $F$-value of approximately $0.70$ and a $p$-value of approximately $0.62$. We view this as indication, that the system at $\tau=1/2$ already reflects an equilibrium configuration well. 
}
\end{center}
\end{figure}

\subsection{Slow compression}
In a simulation protocol, such as described in \secref{sec_simulation}, one needs to guarantee, that the final configuration reflects/is close to a true equilibrium configuration. In practice, this corresponds to the execution of the protocol slowly enough, such that the system does not get quenched into an unfavorable kinetically arrested state.
In particular, we execute the protocol slowly enough, that all configurations throughout the simulation (beyond the initial fast compression given by \equref{eq_compression}) reflect a state close to equilibrium for any instance of parameters across the simulation.

To reinforce this claim, we show simulation results in this section, where we stop the compression and increase of the interaction constants, $U_0$ and $V_0$, after half the number of Monte-Carlo cycles is completed ($\tau=1/2$). 
More specifically, we start the simulation with the same equilibration protocol as described in \secref{sec_simulation}. We abruptly terminate the change of parameters at $\tau=1/2$.
We continue the Monte-Carlo simulation at constant parameters and show that the \textit{foot}-network charge $\Qf$ does not significantly change afterwards. We do this, to illustrate, that the system is close to equilibrium throughout the simulation.
In \figref{fig_Qoftau}, we show distributions of $\Qf$ over $20$ independent simulations as a function of simulation progress $\tau$ (completed number of Monte-Carlo cycles). We denote the average value of $\Qf$ for a constant $\tau$ by $\overline{\mathcal{Q}}_f$. We test whether the $\overline{\mathcal{Q}}_f$ vary significantly after stopping the compression at $\tau=1/2$. To this end, we set up a hypothesis test for the equality of all values with $\tau \geq 1/2$ \cite{miller1997beyond} (the conditions for the hypothesis test: equal variance, approximate normality, independence, were checked).
The null hypothesis is given by $\mathrm{H}_0$: ``All measured $\overline{\mathcal{Q}}_f(\tau \geq 1/2)$ do not differ significantly'' i.e., the true average $\Qf$ from all possible configurations, which can theoretically be obtained in simulation, is constant.
The alternate hypothesis is $\mathrm{H}_a$: ``At least one pair is not equal.''. If the  alternate hypothesis is true, we have to assume that the $\overline{\mathcal{Q}}_f(\tau \geq 1/2)$ vary significantly, and so the simulation has not reached near-equilibrium. Through a one-way $F$-test \cite{2020SciPy-NMeth}, we find $F \approx 0.70$ and a $p$-value of $0.62$.
The value $F$ denotes the variance between $\overline{\mathcal{Q}}_f$ for different $\tau$ divided by the variance for a constant $\tau$. The $p$-value indicates that a distribution of $\overline{\mathcal{Q}}_f$ with such a variance, or more, is expected to occur with probability $p=0.62$. We conclude, that given this data, the null hypothesis $\mathrm{H}_0$ can not be rejected with any reasonable significance level. 
We furthermore infer that, in combination with the absence of a clear up- or downwards trend after $\tau = 1/2$, we have reason to believe, that the observables do not change after stopping compression.
This can be considered indication, that at $\tau = 1/2$, the system resides in a state which reflects equilibrium well.

\section{Network generation }
\label{sec_app_netgen}

The networks, considered in this work, are generated by considering the Delaunay triangulation \cite{aurenhammer2013voronoi} of the union of the geometrical centers of both axes of the L-shaped particles (see~\figref{fig_app_graphgen}, {corresponding to the particle snapshot in \subfigref{fig_network_concept}{a}}). The emerging network is disconnected into the two network species, by deleting edges between opposing {foot} and {leg} vertices (\figref{fig_app_graphgen}, gray). 
Furthermore, to obtain a physical picture of these two intertwined networks as a pair of smectic-like layers with a topological charge conservation, we require two further systematic modifications. Firstly,
we assign the boundary to the \textit{leg}-network.
This corresponds to assuming that the \textit{leg} of the particles align preferably parallel with the wall, while other local configurations are interpreted as a defect.
Such a presumed uniform alignment rule stands at the basis of any topological conservation law in confinement.
Secondly, to ensure that the final networks represent global layer structures, 
we transform the triangular meshes into simple lines through merging vertices that form empty triangles (without changing the hierarchy between the two networks). 
Thereby we delete any  empty loops which are not compatible with the concept of alternating layers required for topological charge conservation. 
This is in accordance with the layer and half-layer picture of conventional smectics and can be readily applied to uniaxial rods by considering the limit $\Lb=0$. The final network, after applying this protocol to \figref{fig_app_graphgen}, is shown in \subfigref{fig_network_concept}{c}.\\

\begin{figure}[t]
\begin{center}
\vspace{0.5cm}
\includegraphics[width=0.85\linewidth]{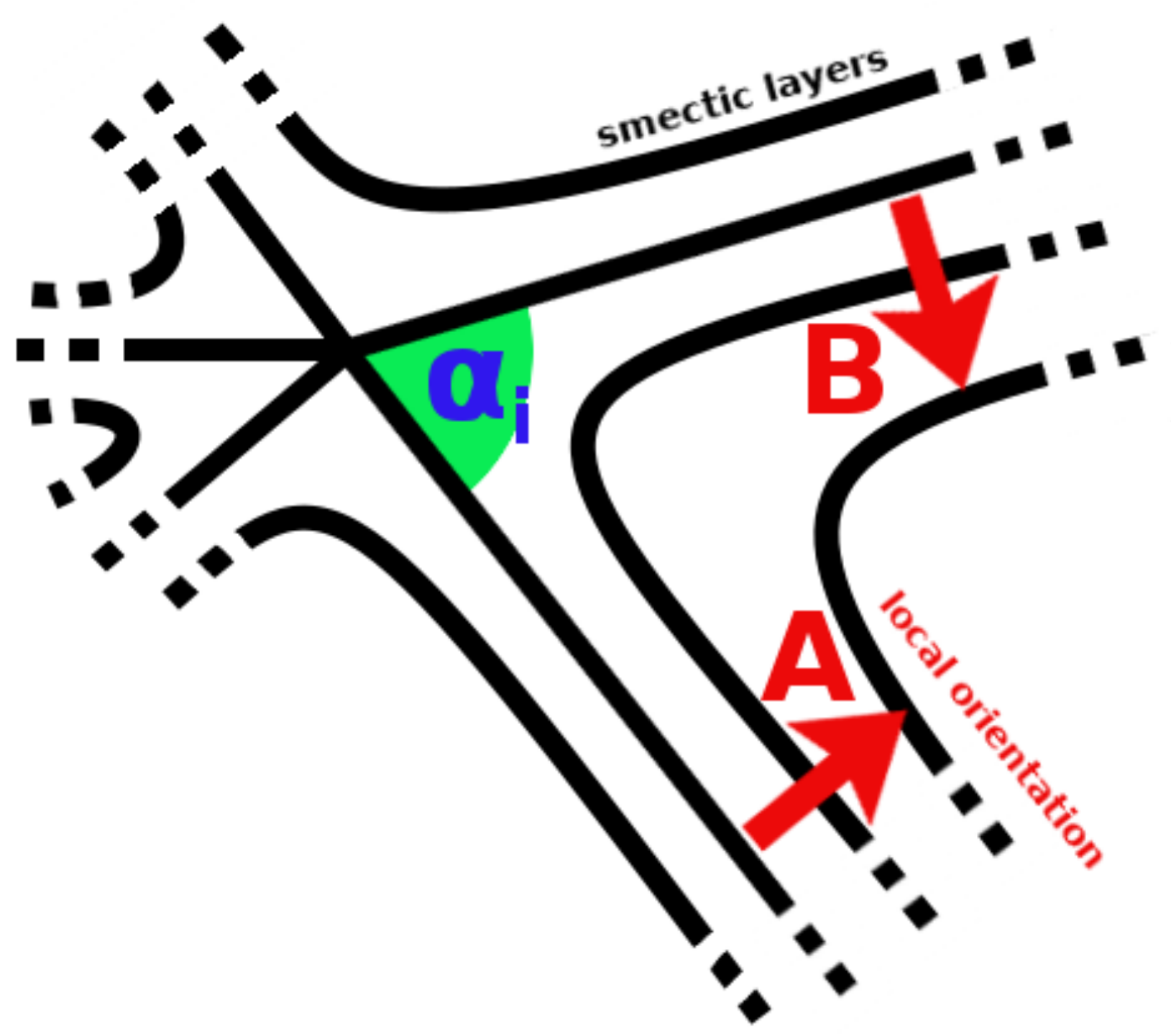}
\caption{\label{fig_app_smecdef} Schematic of an orientational topological defect in a smectic liquid crystal. The defect is a point, where the local orientation $\nofr$, typically orthogonal to the smectic layers, indicated as black lines, is ill-defined. The rotation $\Delta\phi_i$ from $\mathbf{A}$ to $\mathbf{B}$ is equal to $\alpha_i-\pi$. Therefore the total rotation $\Phi$ is equal to $\sum_i^n \Delta \phi_i = 2\pi(1-n/2)$, with the number of outgoing layers $n$. The topological charge is equal to $\Phi/2\pi = 1-n/2$, in analogy to \equref{eq_netwQ}.} 
\end{center}
\end{figure}

\section{Analogy to topological charges in smectics \label{sec_app_convsmec}}

The network topological charges, introduced in Sec.~\ref{sec_nettop} share a close relation with the orientational topological charges, typically considered in smectic liquid crystals \cite{kamien2016topology,chen2009symmetry,exp_studies}. 
In this manuscript, we assign a network topological charge $q = 1- \degr/2$ to any vertex in the observed network via its degree $\degr$ (see \equref{eq_netwQ}). Similarly, topological charges of defects in smectic systems can be understood in terms of adjacent layers \cite{machon2019,aharoni2017composite}.\\

\begin{figure}[t]
\begin{center}
\vspace{0.5cm}
\includegraphics[width=1.0\linewidth]{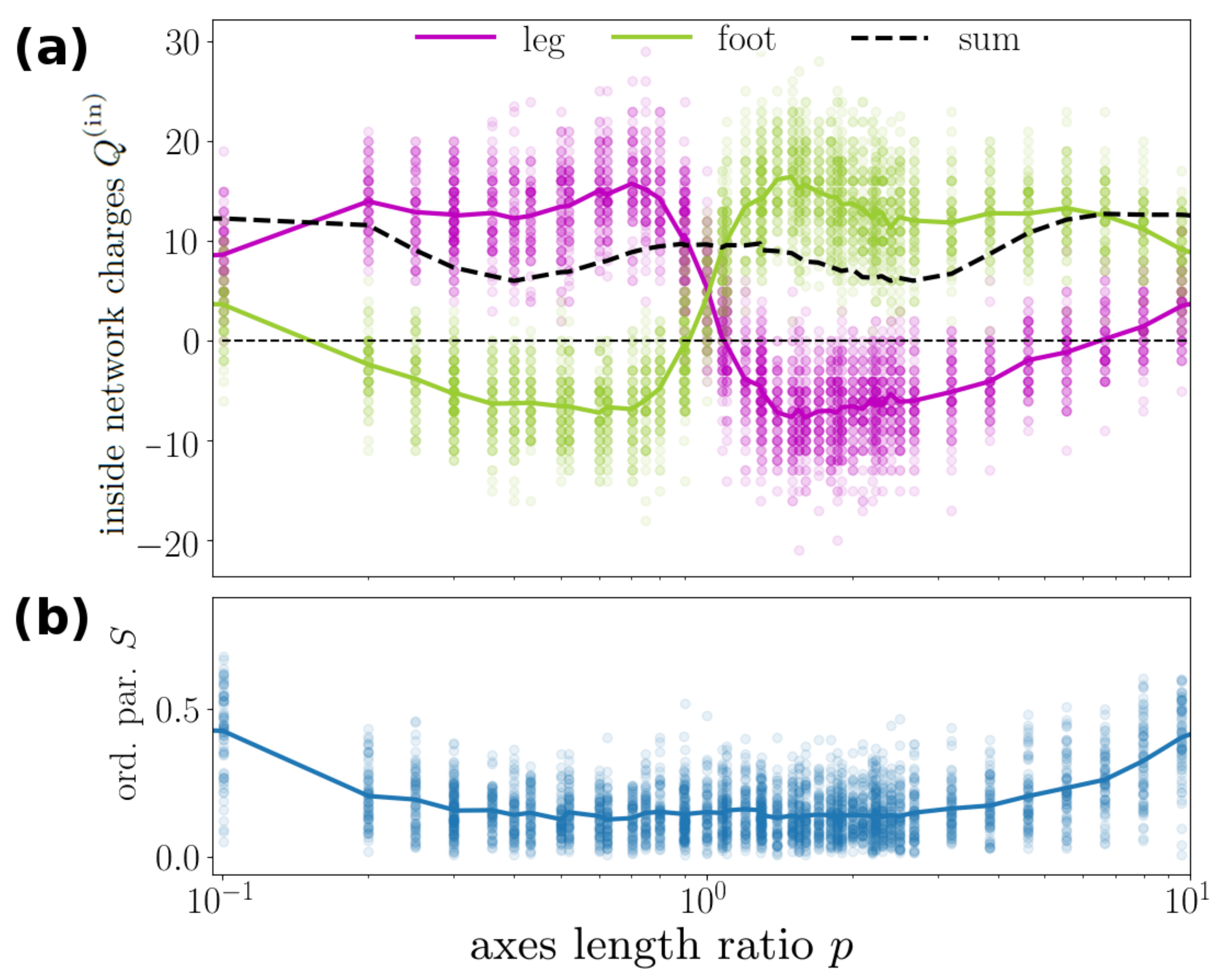}
\caption{\label{fig_app_Lratlonger}
Simulation results of liquid crystals composed of L-shaped particles for a range of different length ratios $p$ of both particle axes as \figref{fig_Qofrat} in the main text. Through the logarithmic scaling of the horizontal axis, the symmetry around $p=1$ becomes apparent.
} 
\end{center}
\end{figure}

Orientational topological defects in liquid crystals, that display local alignment of the molecules, can be understood as singular locations in space, where the local orientation $\nofr$ undergoes a discontinuous jump and is therefore ill-defined. In smectic systems, where the particles additionally arrange in layers, this can happen across grain boundaries in two and three dimensions, or across points in two dimensions. A schematic of a two-dimensional point defect is depicted in \figref{fig_app_smecdef}. Smectic layers are represented by black lines.
In this particular example, five layers meet in a singular point. Around this point, the layers, and thus $\nofr$, typically at a constant angle to the layers, are continuously bent.
The strength of the defect is characterized by the total rotation of $\nofr$ traversing the defect in counterclockwise direction. Consider one wedge of the rotation $\mathbf{A}\rightarrow\mathbf{B}$ between two layers at an angle $\alpha_i$: The rotation angle is equal to $\Delta \phi_i = \alpha_i - \pi$. Consequently, the total rotation around the defect is equal to 
\begin{equation}
    \Phi = \sum_i^n \Delta \phi_i = 2\pi(1-n/2),
\end{equation}
with the number of outgoing layers $n$. The topological charge of the defect is defined by $\Phi/2\pi$ resulting in \equref{eq_netwQ}, with the vertex degree $\degr$ identified with $n$.\\

\section{Larger axes length ratios \label{sec_app_Lratlonger}}
We denote the conventionally short (length $\Lf$) horizontal axis of the letter L by \textit{foot} and the conventionally long (length $\Lb$) vertical axis by \textit{leg}. The length ratio $p$ of the axes can in practice vary between $0$ and $\infty$. If the symmetry between the two axes is not broken, e.g., by assignment of boundary charges to any of the two corresponding network species, the physical observables should generally be symmetric around $p=1$.
This is confirmed in \figref{fig_app_Lratlonger}, where \textbf{(a)} the inside network charges and \textbf{(b)} the global orientational order parameter $S$ are depicted. The figure features the simulation data depicted in \figref{fig_Qofrat} supplemented with simulation data for larger $p$. 
In particular, we verify that $S(p) = S(1/p)$, as well as $\Qbb(p) = \Qfb(1/p)$, i.e.,
the symmetry around $p=1$ is clearly visible.

\end{document}